\documentclass[12pt,a4paper]{article}
\usepackage{indentfirst}
\usepackage{epsfig}

\begin{document}

\begin{center}
  {\Large \bf  System Size and Centrality Dependence }
\end{center}
\begin{center}
  {\Large \bf of the Balance Function in A + A }
\end{center}
\begin{center}
  {\Large \bf Collisions at $\sqrt{s_{NN}} = 17.2$ GeV}
\end{center}

\vspace{1.0cm}

\begin{center}
  {\bf The NA49 Collaboration}
\end{center}

\vspace{1.0cm}

\begin{abstract}

Electric charge correlations were studied for p+p, C+C, Si+Si and 
centrality selected Pb+Pb collisions at $\sqrt{s_{NN}} = 17.2$ GeV 
with the NA49 large acceptance detector at the CERN-SPS. In particular, 
long range pseudo-rapidity correlations of oppositely charged particles 
were measured using the Balance Function method. The width of the Balance 
Function decreases with increasing system size and centrality of the 
reactions. This decrease could be related to an increasing delay of 
hadronization in central Pb+Pb collisions.

\end{abstract}

\newpage

\begin{center}
  {\bf The NA49 Collaboration}
\end{center}

\vspace{0.5cm}
\noindent
C.~Alt$^{9}$, T.~Anticic$^{21}$, B.~Baatar$^{8}$,D.~Barna$^{4}$,
J.~Bartke$^{6}$, 
L.~Betev$^{9,10}$, H.~Bia{\l}\-kowska$^{19}$, A.~Billmeier$^{9}$,
C.~Blume$^{9}$,  B.~Boimska$^{19}$
% ...

\vspace{0.5cm}
\noindent
% ...
$^{4}$KFKI Research Institute for Particle and Nuclear Physics, Budapest, Hungary.\\
% ...
$^{6}$Institute of Nuclear Physics, Cracow, Poland.\\
% ...
$^{8}$Joint Institute for Nuclear Research, Dubna, Russia.\\
$^{9}$Fachbereich Physik der Universit\"{a}t, Frankfurt, Germany.\\
$^{10}$CERN, Geneva, Switzerland.\\
% ...
$^{19}$Institute for Nuclear Studies, Warsaw, Poland.\\
% ...
$^{21}$Rudjer Boskovic Institute, Zagreb, Croatia.\\

\section{Introduction}
Collisions of heavy ions have been used throughout the last decades in 
order to investigate the possible formation of the quark-gluon plasma (QGP) 
\cite{QGP}, by studying a variety of characteristics \cite{QM}. At the early 
stage of these collisions an extended region with large energy density may be 
produced, where hadronic may be replaced by quark-gluon degrees of freedom 
possibly leading to a new partonic phase of matter. In the subsequent evolution, 
the system dilutes and cools down, hadronizes and finally decays into free 
hadrons. These final state hadrons carry only indirect information about the 
early stage of the collision.

Numerous observables, such as particle yields and measures of correlations and 
fluctuations, have been proposed that could signal the possible transition 
from the hadronic to the partonic phase. Recent data suggest that conditions, 
consistent with the occurrence of the QCD phase transition, are indeed reached 
\cite{Alb94} in Pb+Pb collisions at 158\emph{A} GeV at the CERN SPS. Moreover, 
results from the study of the energy dependence of single particle yields and 
spectra, suggest that a deconfined phase starts to be formed in the early stage 
of the reaction at low SPS energies \cite{QM04}. The study of correlations and 
fluctuations is expected to provide additional information on the reaction 
mechanism of high energy nuclear collisions. In particular, event-by-event charge 
and mean $p_T$ fluctuations, have been analyzed over the past years \cite{Stock99}. 
Another important measure of correlations, the Balance Function (BF), was 
introduced by Bass, Danielewicz and Pratt \cite{Pratt}. It measures the 
correlation of the oppositely charged particles produced during a heavy ion 
collision and its width can be related to the time of hadronization. The BF is 
derived from the charge correlation function that was used to study the 
hadronization of jets in p+p collisions at the ISR \cite{Drij} and $e^-+e^+$ 
annihilations at PETRA \cite{Aih}. The first results on the BF were obtained for 
Au+Au collisions by the STAR collaboration at RHIC \cite{STAR}.

In this paper we study the BF in p+p, C+C, Si+Si and centrality selected Pb+Pb 
collisions at a beam energy of 158\emph{A} GeV, corresponding to a center-of-mass 
energy of $\sqrt{s_{NN}} = 17.2$ GeV per nucleon pair. The data were obtained with 
the NA49 detector at the CERN SPS.

\section{The Balance Function Method}

The motivation for studying the Balance Function comes from the idea that hadrons 
are produced locally as oppositely charged particle pairs. Particles of such a pair 
are separated in rapidity due to the initial momentum difference and secondary 
interactions with other particles.

Particles of a pair that was created earlier are separated further in rapidity 
because of the expected large initial momentum difference and the long lasting 
rescattering phase. On the other hand, oppositely charged particle pairs that 
were created later are correlated within a smaller interval $\Delta y$ of the 
relative rapidity. Our aim is to measure the degree of this separation of the 
balancing charges and to find possible indications for delayed hadronization.

In this paper the BF is used in order to examine the pseudo-rapidity ($\eta$) 
correlation of charged particles. It is defined as a difference of the correlation 
function of oppositely charged particles and the correlation function of like-charge 
particles normalized to the total number of particles. The general definition of the 
BF reads \cite{Pratt}:

\[B(P_2|P_1) = \frac{1}{2} \Big[ \frac{N(b,P_2|a,P_1) -
N(a,P_2|a,P_1)}{N(a,P_1)} + \]

\begin{center}
\begin{equation}
\frac{N(a,P_2|b,P_1) - N(b,P_2|b,P_1)}{N(b,P_1)} \Big], \label{BF_DEF}
\end{equation}
\end{center}

\noindent where $a$ and $b$ could be different kinds of particles, whereas $P_1$ 
and $P_2$ could be intervals in pseudo-rapidity. For example $a$ could refer to 
all negative particles and $b$ to all positive particles. Alternatively $P_2$ 
could be an interval of the relative pseudo-rapidity $\Delta \eta = |\eta_b - \eta_a|$ 
of the oppositely charged particles, whereas $P_1$ could be the interval of the 
pseudo-rapidity of the produced particles that is covered by the detector. In the 
numerator, $N(b,P_2|a,P_1)$ represents a conditional probability of observing a 
particle of type $b$ in bin $P_2$ given the existence of a particle of type $a$ 
in bin $P_1$. The terms $N(b,P_2|a,P_1)$, $N(a,P_2|a,P_1)$, $N(a,P_2|b,P_1)$ and
$N(b,P_2|b,P_1)$ are calculated using pairs from each event and the resulting 
values are summed over all events. For example, the term $N(b,P_2|a,P_1)$ is 
calculated by counting all possible combinations of a positive particle in $P_2$ 
and a negative particle in $P_1$ in an event and summing the number of combinations 
over all events. The other three terms are calculated analogously. The terms
$N(a,P_1)$ and $N(b,P_1)$ are the total number of negative and positive particles, 
respectively, that are within the studied pseudo-rapidity interval $P_1$, summed 
over all events.

In our case, $a$ and $b$ are the negative and positive particles respectively that 
are within the pseudo-rapidity interval $P_1$ and have a pseudo-rapidity difference 
$\Delta \eta$. So the definition of the BF takes the following form:

\begin{center}
\begin{equation}
B(\Delta \eta) = \frac{1}{2} \Big[ \frac{N_{+-}(\Delta \eta) -
N_{--}(\Delta \eta)}{N_{-}} + \frac{N_{-+}(\Delta \eta) -
N_{++}(\Delta \eta)}{N_{+}}  \Big].
\label{BF_DEF2}
\end{equation}
\end{center}

\vspace{0.5 cm}

The most interesting property of the BF is its width. Early stage hadronization is 
expected to result in a broad BF, while late stage hadronization leads to a narrower 
distribution \cite{Pratt}. The width of the BF can be characterized by the weighted 
average $\langle \Delta \eta \rangle$:

\begin{center}
\begin{equation}
\langle \Delta \eta \rangle = \sum_{i=0}^k{(B_i \cdot \Delta \eta _i)}/\sum_{i=0}^k{B_i},
\label{width}
\end{equation}
\end{center}

\noindent where \emph{i} is the bin number of the BF histogram.

\section{Experimental Setup}
\begin{figure}[ht]
\begin{center}
\epsfig{angle=0,file= 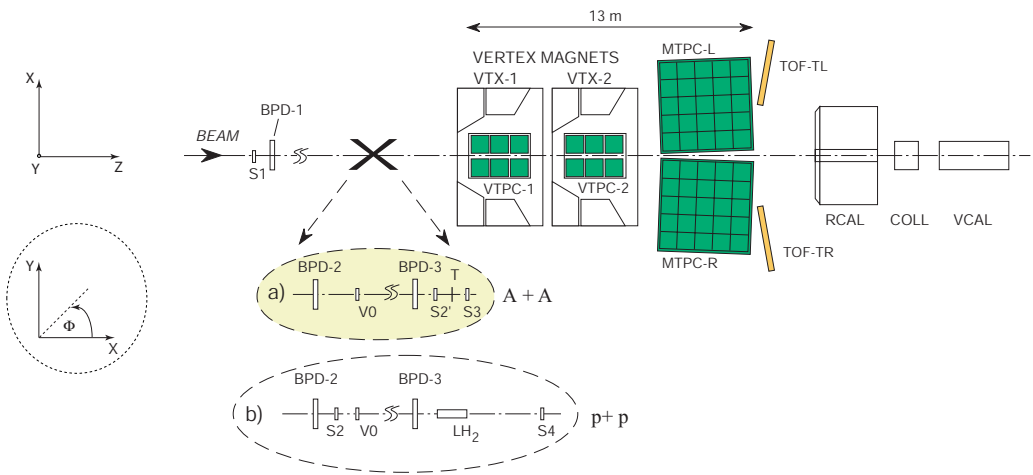,width=13cm}
\end{center}
\caption{The experimental set--up of the NA49 experiment with
different beam definitions and target arrangements (color online).}
\label{na49_setup}
\end{figure}

The NA49 detector \cite{na49_nim} is a wide acceptance hadron
spectrometer for the study of hadron production in collisions of
hadrons or heavy ions at the CERN SPS. The main components 
are four large-volume Time Projection Chambers (TPCs)
(Fig. \ref{na49_setup}) which are capable of detecting 80\% of
some 1500 charged particles created in a central Pb+Pb collision
at 158\emph{A} GeV. Two chambers, the Vertex TPCs (VTPC-1 and
VTPC-2), are located in the magnetic field of two super-conducting
dipole magnets (1.5 and 1.1 T, respectively), while the two others
(MTPC-L and MTPC-R) are positioned downstream of the magnets
symmetrically to the beam line. The set--up is supplemented by two
Time of Flight (TOF) detector arrays and a set of calorimeters.
The data presented in this paper are analyzed with a global
tracking scheme \cite{na49_global}, which combines track segments
that belong to the same physical particle but were detected in
different TPCs. The NA49 TPCs allow precise measurements of
particle momenta $p$ with a resolution of $\sigma(p)/p^2 \cong
(0.3-7)\cdot10^{-4}$ (GeV/c)$^{-1}$.

The targets are C (561 mg/cm$^{2}$), Si (1170 mg/cm$^{2}$) disks 
and a Pb (224 mg/cm$^{2}$) foil for ion collisions and a liquid hydrogen
cylinder (length 20 cm) for hadron interactions. They are positioned 
about 80 cm upstream from VTPC-1.

Pb beam particles are identified by means of their charge as seen
by a Helium Gas-Cherenkov counter (S2') and proton beam particles 
by a 2 mm thick scintillator (S2). Both detectors are situated in 
front of the target. For p, C and Si beams, interactions in the 
target are selected by anti-coincidence of the incoming beam particle 
with a small scintillation counter (S4) placed on the beam line 
between the two vertex magnets. For p+p interactions at 158 GeV 
this counter selects a (trigger) cross section of 28.5 mb out of 31.6
mb of the total inelastic cross section. For Pb-ion beams, an
interaction trigger is provided by anti-coincidence with a Helium
Gas-Cherenkov counter (S3) directly behind the target. The S3
counter is used to select minimum bias collisions by requiring a
reduction of the Cherenkov signal by a factor of about 6. Since
the Cherenkov signal is proportional to $Z^2$, this requirement
ensures that the Pb projectile has interacted with a minimal
constraint on the type of interaction. This setup limits the
triggers on non-target interactions to rare beam-gas collisions,
the fraction of which proved to be small after cuts, even in the
case of peripheral Pb+Pb collisions.

The centrality of a collision is selected (on-line for central Pb+Pb, 
Si+Si and C+C and off-line for minimum bias Pb+Pb interactions) by a 
trigger using information from a downstream calorimeter (VCAL), which 
measures the energy $E_0$ of the projectile spectator nucleons.

\section{Data Analysis}

\subsection{Data Sets}

The data sets used in this analysis come from p+p, C+C, Si+Si and Pb+Pb collisions 
at 158\emph{A} GeV. For Pb+Pb interactions data with both central ($2 \cdot 10^5$) 
and minimum bias trigger ($6 \cdot 10^5$) have been analyzed in order to study the 
centrality dependence of the BF. The minimum bias data were subdivided into six 
different centrality classes \cite{Cooper} according to the energy recorded by the 
VCAL, from class Veto 1 (the most central collisions) to class Veto 6 (the most 
peripheral collisions). The most central Pb+Pb interactions correspond to $5\%$ of 
the total geometric cross section (Table \ref{data_sets}). Since minimum bias data 
provide only a small number of central collisions, we used in addition trigger 
selected central data. Finally, we analyzed three different data sets (Table \ref{sets}) 
of Pb+Pb minimum bias events coming from two different data-taking periods ($1996$ - Data 
Set 1 and Data Set 2,  $2000$ - Data Set 3) with opposite magnetic field polarities 
(positive field polarity - Data Set 2 and Data Set 3, negative field polarity - 
Data Set 1)  in order to estimate the systematic uncertainties (Table \ref{sets}).

The event centrality is characterized by the mean impact parameter $\langle b \rangle$ 
and the corresponding number of wounded nucleons $\langle N_W \rangle$. For each bin 
of centrality these quantities were determined by use of the Glauber model as 
implemented in the VENUS event generator \cite{venus_ref}. In order to estimate the 
correlation between the energy deposited in the VCAL and $\langle b \rangle$ or 
$\langle N_W \rangle$ minimum bias VENUS events were processed through the GEANT 
detector simulation code, and the energy deposited in the VCAL was simulated. All 
these quantities are listed in Table \ref{data_sets}.

\subsection{Event and Track Selection}

In order to reduce the contamination from non-target events and non-vertex tracks, 
selection criteria were imposed both at the event and the track level.

Events were selected that had a proper position of the reconstructed primary vertex. 
The vertex coordinate $V_z$ along the beam axis had to fulfill $|V_z - V_{z_0}| < \Delta z$ 
where the values of the central position $V_{z_0}$ and the range $\Delta z$ are shown 
in Table \ref{sets} for p+p, C+C, Si+Si and Pb+Pb reactions, respectively. In addition 
the vertex coordinates $V_x$ and $V_y$ perpendicular to the beam axis had to fulfill 
$|V_x - V_{x_0}| < \Delta x$ and $|V_y - V_{y_0}| < \Delta y$, where the values 
$V_{x_0}, V_{y_0}$ and $\Delta x$, $\Delta y$ can also been seen in Table \ref{sets} 
for all the data samples analyzed.

Selection criteria at the track level were imposed in order to reduce the contamination 
by tracks from weak decays, secondary interactions other sources of non vertex tracks. 
Thus, an accepted track had to have an extrapolated distance of closest approach $d_x$ 
and $d_y$ of the particle at the vertex plane within the range: $|d_x| < 2.0$ cm and 
$|d_y| < 1.0$ cm. In addition the potential number of points in the detector for the 
selected tracks had to be more than 30. To suppress double counting due to track splitting, 
the ratio of the number of reconstructed points to the potential number of points was 
required to be larger than $0.5$.

\begin{figure}[ht]
\begin{center}
\epsfig{angle=0,file=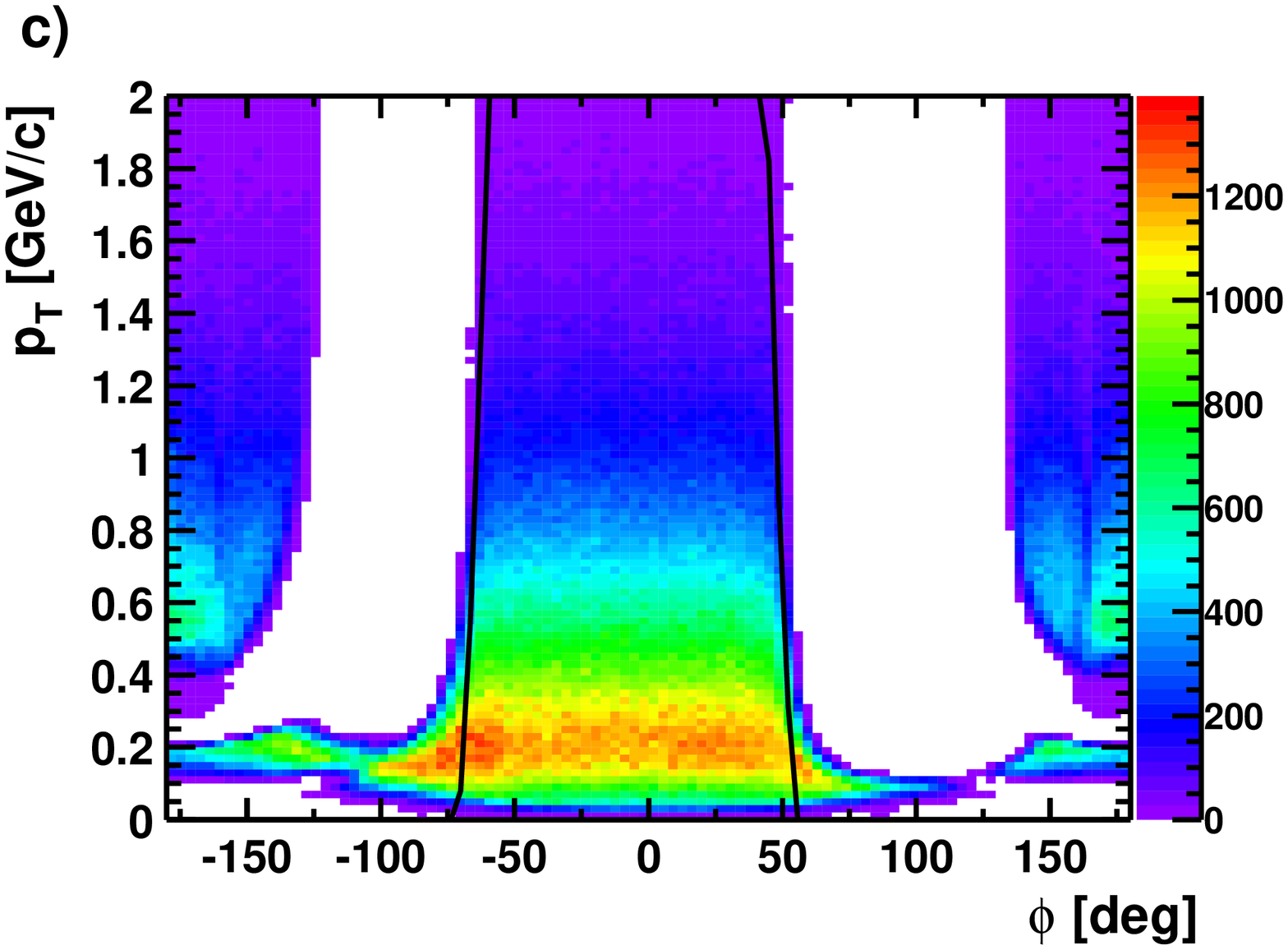,width=6cm,height=5cm}
\epsfig{angle=0,file=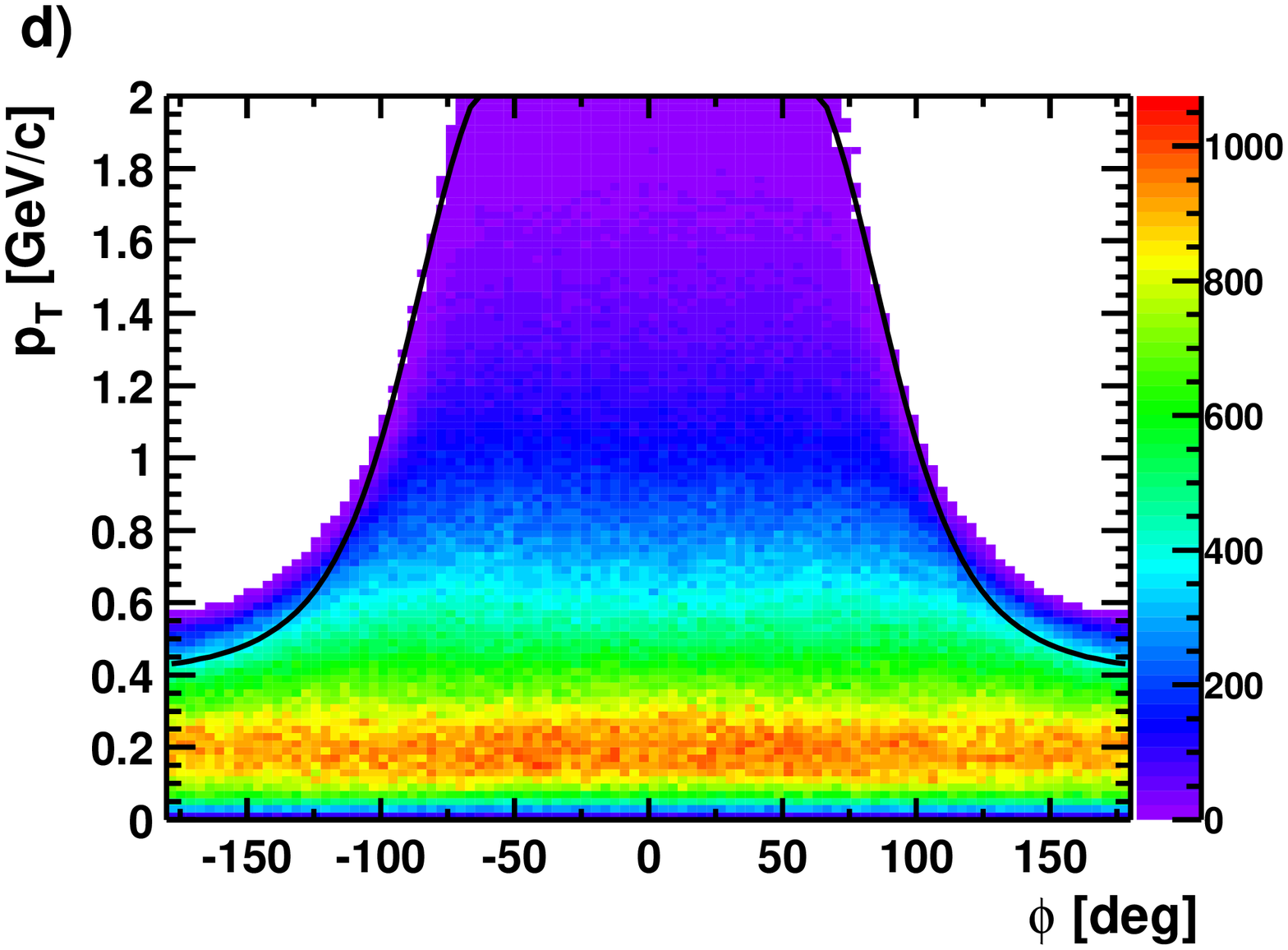,width=6cm,height=5cm}
\end{center}
\caption{The acceptance curves in the $p_{T} - \phi$ plane
for $2.5 < y < 2.7$ (left plot) and $4.1 < y < 4.3$ (right plot)
at $\sqrt{s_{NN}} = 17.2$ GeV (color online).} \label{Acc2}
\end{figure}

\begin{figure}[ht]
\begin{center}
\epsfig{angle=270,file=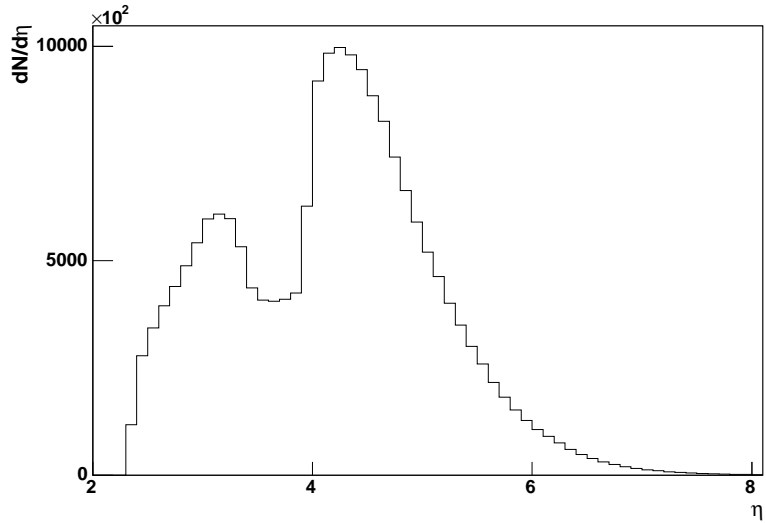,width=10cm}
\end{center}
\caption{The pseudo-rapidity distribution of the accepted charged particles
in central Pb+Pb collisions at $\sqrt{s_{NN}} = 17.2$ GeV.} \label{eta}
\end{figure}

The NA49 detectors provide large acceptance in momentum space; however the acceptance 
in the azimuthal angle $\phi$ is not complete. The boundary of the acceptance region 
can be described with the formula \cite{Jacek}:

\begin{equation}
p_{T}(\phi) = \frac{1}{A+(\frac{D+\phi}{C})^{6}}+B,
\label{Acc_Jac}
\end{equation}

\noindent where the values of the parameters $A$, $B$, $C$ and $D$ depend on the
rapidity interval and are given in Table \ref{abcd} (see Fig. \ref{Acc2} for examples). 
The inclusive pseudo-rapidity distribution after applying the acceptance filter can 
be seen in Fig. \ref{eta}.

Finally we required tracks to additionally satisfy the following criteria: 
$0.005 < p_T < 1.5$ GeV/c and $ 2.6 < \eta < 5.0$. As shown in Fig. \ref{eta} 
the phase space analyzed covers most of the forward rapidity region, where the 
geometric acceptance is maximal.

\subsection{Results}

In this section, we will present results on the BF (Eq. \ref{BF_DEF2}) measured in 
p+p, C+C, Si+Si and Pb+Pb at $\sqrt{s_{NN}} = 17.2$ GeV that were subjected to the event 
and track quality as well as to the phase space cuts described in the previous section.

In order to study the centrality dependence of the BF, we analyzed Pb+Pb collisions 
that were divided into six centrality (Veto) classes \cite{Cooper}, from 1 (the most 
central collisions) to 6 (the most peripheral ones) (Table \ref{data_sets}).

\begin{figure}[ht]
\begin{center}
\epsfig{angle=0,file=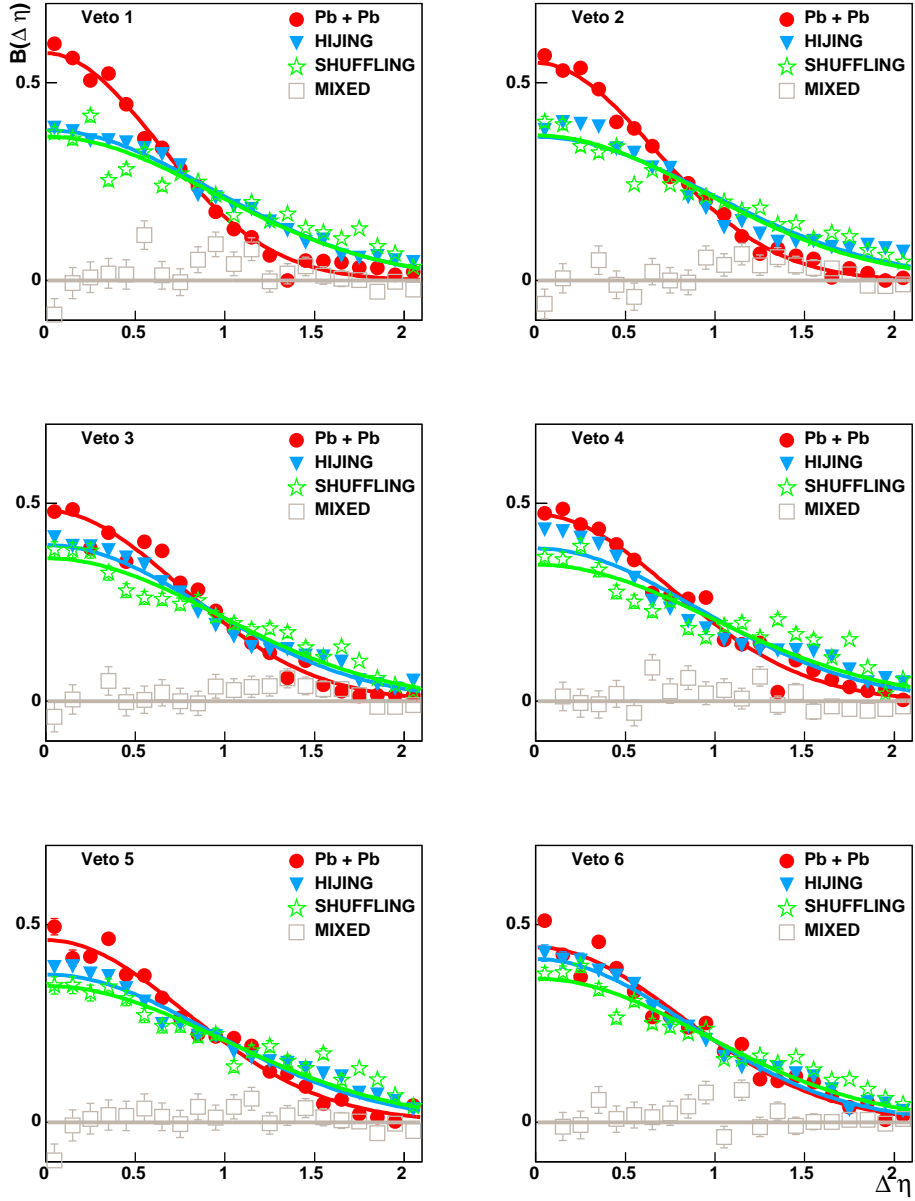,width=12cm}
\end{center}
\caption{The BF versus $\Delta \eta$ for different centrality classes of
Pb+Pb collisions for real data as well as for shuffled, mixed and HIJING 
events. The curves show Gaussian fits (color online).} \label{PbPb}
\end{figure}

The results are shown in Fig. \ref{PbPb} where the BF is plotted as a function of 
$\Delta \eta$, the pseudo-rapidity difference of the charged particles. The error on 
each measured point is the statistical error. For visual comparisons the distributions 
were fitted with a Gaussian function having a fixed mean at zero (curves in Fig. \ref{PbPb}). 
From inspection of Fig. \ref{PbPb} as well as from the values of the weighted 
average $\langle \Delta \eta \rangle$ that are listed in Table \ref{sigma_results},
we notice that the width $\langle \Delta \eta \rangle$ of the BF is narrower for the most 
central collisions (Veto 1) than for the peripheral ones (Veto 6). It should be mentioned, 
that for the calculation of the width (Eq. \ref{width}) we excluded the first 
point of each distribution, since from \cite{Distortions} it was shown that this point is 
significantly influenced by Coulomb interactions and Bose-Einstein correlations.

\begin{figure}[ht]
\begin{center}
\epsfig{angle=0,file=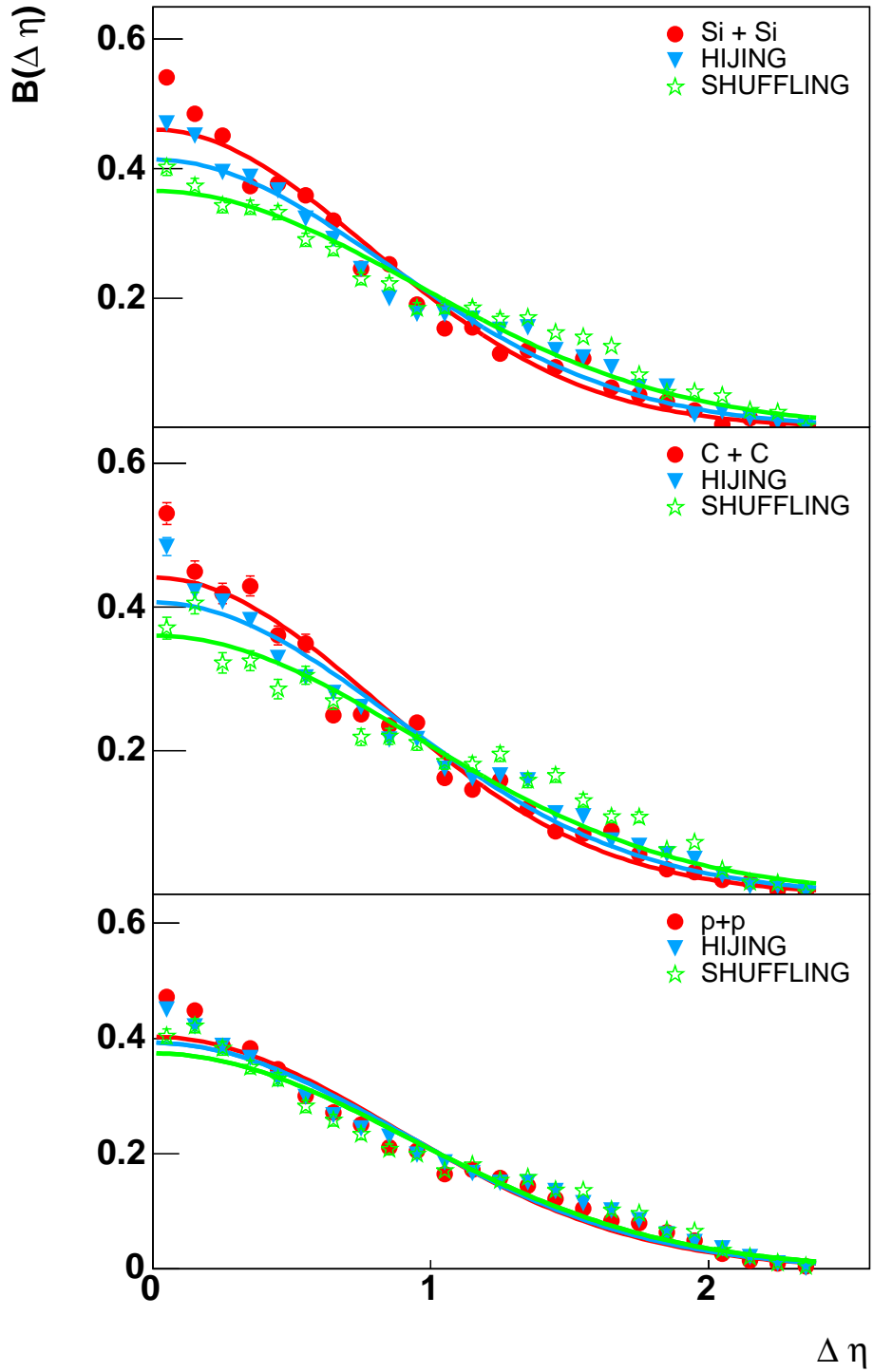,width=12cm}
\end{center}
\caption{The BF versus $\Delta \eta$ for real, shuffled and HIJING 
events (together with the Gaussian fits) for Si+Si (upper panel), 
C+C (middle panel) and p+p (lower panel) collisions (color online).} \label{Lighter}
\end{figure}

Furthermore, in order to extend the method to a system size study, we have analyzed 
C+C and Si+Si collisions at $\sqrt{s_{NN}} = 17.2$ GeV. The BFs for the data samples 
of these two systems are shown in Fig. \ref{Lighter}. The distributions are wider than 
those of the most central Pb+Pb collisions and tend to be similar to the ones coming 
from the most peripheral Pb+Pb interactions (Veto 6). This conclusion is confirmed by 
the corresponding values of $\langle \Delta \eta \rangle$ displayed in Table \ref{sigma_results}.

In addition, we have studied p+p interactions at $\sqrt{s_{NN}} = 17.2$ GeV. The 
resulting BF distribution shown in Fig. \ref{Lighter} is significantly wider than that 
for Pb+Pb interactions. The calculated widths $\langle \Delta \eta \rangle$ for p+p, 
C+C, Si+Si and all centrality classes of Pb+Pb interactions are summarized in Table 
\ref{sigma_results} along with their statistical errors.

\subsection{Systematic errors}

The systematic errors of the width of the BF were estimated by varying the cuts in 
$V_z$, $d_x$ and $d_y$ and by comparing results obtained from different data taking 
periods. The results are described in this section.

\begin{figure}[ht]
\begin{center}
\epsfig{angle=0,file=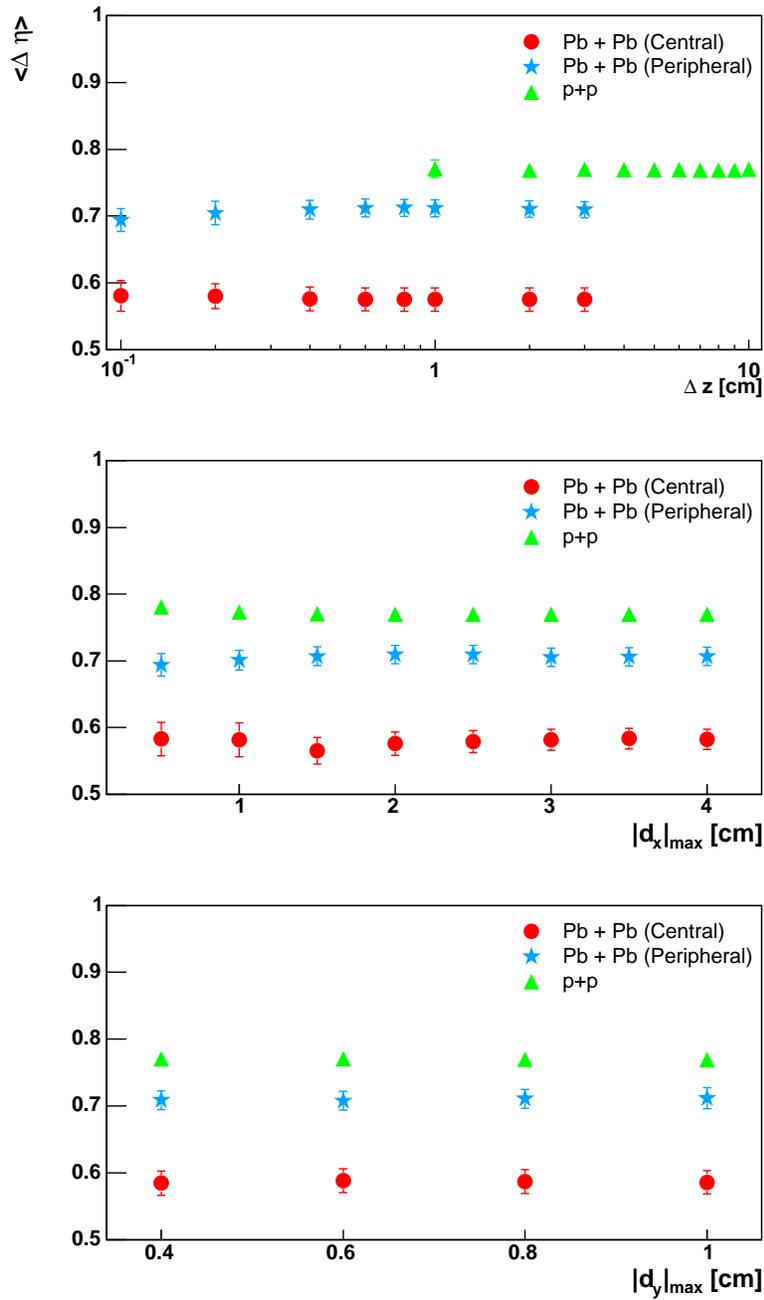,width=10cm}
\end{center}
\caption{The width of the BF as a function of the event selections cut $\Delta z$ 
(upper panel) and track selection cuts $|d_x|_{max}$ (middle panel) and $|d_y|_{max}$ 
(lower panel) for p+p, Pb+Pb peripheral and Pb+Pb central data (color online).} 
\label{systematic}
\end{figure}

The dependence of the width of the BF on the cut $\Delta z$ for the event vertex 
position and the upper limit cuts on the impact parameters $|d_x|$ and $|d_y|$ are 
shown in Fig. \ref{systematic} for p+p and Pb+Pb (central and peripheral) collisions. 
The resulting variations of the width of the BF are used to estimate the systematic 
errors due to contamination of non-target interactions and non-vertex tracks. They 
amount to no more than $0.006$, $0.009$ and $0.003$ for p+p, Pb+Pb peripheral and 
Pb+Pb central collisions, respectively.

Finally, as mentioned in a previous section, we analyzed three different data sets of 
minimum bias Pb+Pb collisions. The observed difference in the BF width are smaller than:
$0.005$, $0.009$, $0.006$ and $0.004$ for Veto 6, Veto 5, Veto 4 and Veto 3 centrality 
selection respectively.

To summarize, the estimated systematic errors of the width of the BF for p+p, C+C, Si+Si, 
Pb+Pb peripheral and Pb+Pb central collisions are no more than : $\pm 0.006$, $\pm 0.010$, 
$\pm 0.012$, $\pm 0.009$ and $\pm 0.003$, respectively.

\section{Discussion}
In this section the results presented in the previous ones, will be compared to models and 
to results from RHIC obtained by the STAR collaboration \cite{STAR}.

The BF for each centrality class was calculated for mixed events that were produced by 
randomly choosing particles from different events with similar vertex position and 
multiplicity. As shown in Fig.\ref{PbPb}, the BF for mixed events goes to zero because of 
the removal of correlations caused by global charge conservation. Another method of mixing 
was applied to the data sample in order to estimate the maximum possible value of the width 
of the BF while retaining the constraint of charge conservation. This shuffling procedure 
\cite{STAR} is a mixing method in which the value of the pseudo-rapidity of each track is 
taken randomly from the collection of pseudo-rapidity values of the tracks in the same event, 
whilst keeping the charge of each track the same. The BF for shuffled data is broader for each 
centrality class than the one obtained from the real data (Fig. \ref{PbPb}). The values of 
$\langle \Delta \eta \rangle$ for the shuffled data analysis are listed in 
Table \ref{sigma_results_shuf}.

Finally, in order to further investigate the origin of the system size and centrality 
dependence of the BF, we generated p+p, C+C and Si+Si collisions as well as centrality 
selected  Pb+Pb interactions at $\sqrt{s_{NN}} = 17.2$ GeV using the HIJING event generator 
\cite{Hijing}. The model is based on the excitation of strings and their subsequent 
hadronization according to the LUND model. The latter contains short range correlations of 
oppositely charged hadrons which are consistent with measurements from $e^{+}+e^-$ annihilations. 
The rescattering of produced hadrons is not included in the model.

The generated data sets were analyzed with and without applying the NA49 acceptance filter. 
The results revealed that the acceptance filter slightly increases the width by about $4 \%$. 
This suggest that this filter removes a fraction of balancing charges. The filtered 
distributions for Pb+Pb collisions and interactions of lighter systems are plotted in 
Fig. \ref{PbPb} and Fig. \ref{Lighter} respectively. The values of the widths are 
included in Table \ref{sigma_results_shuf}. The BF for HIJING is independent of centrality 
and system size and is wider than the one calculated from the real data for central, 
mid-central and mid-peripheral collisions. On the other hand both HIJING and real data 
distributions tend to be similar for the most peripheral Pb+Pb collisions (Veto 6) as well 
as for the lighter systems.

\begin{figure}[ht]
\begin{center}
\epsfig{angle=270,file=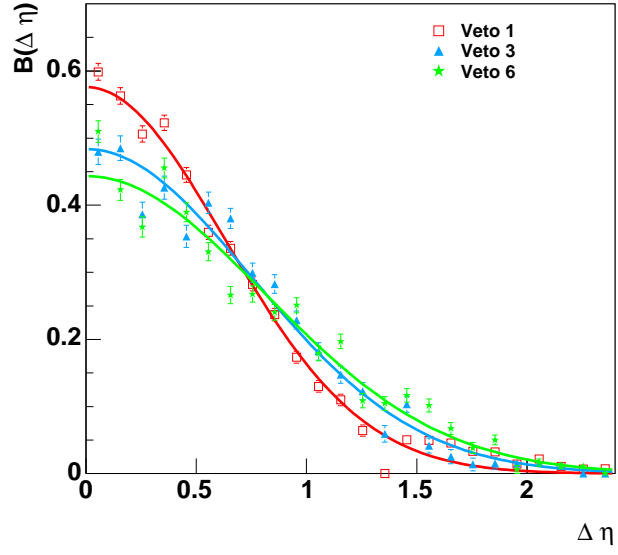,width=8cm}
\end{center}
\caption{The BF versus $\Delta \eta$ for three centrality
classes of Pb+Pb collisions together with the Gaussian fits. The
distributions were normalized to the same integral for this comparison (color online).}
\label{b_same}
\end{figure}

In order to demonstrate the dependence of the BF 's width $\langle \Delta \eta \rangle$
on the centrality class in Pb+Pb interactions, the BFs in different centrality bins were 
normalized to the same area and plotted on the same graph (Fig. \ref{b_same}).
A significant narrowing of the BF width with increasing centrality is observed.

\begin{figure}[ht]
\begin{center}
\epsfig{angle=270,file=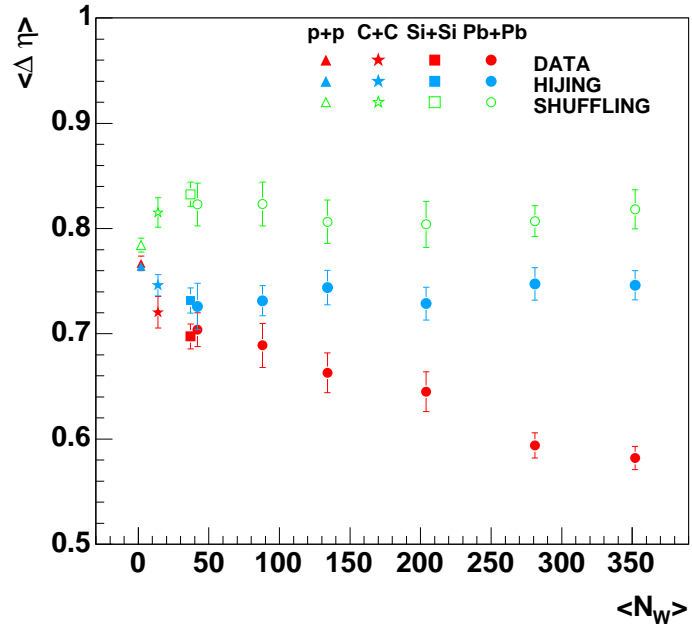,width=9cm}
\end{center}
\caption{The dependence of the BF 's width on the number of
wounded nucleons for p+p, C+C, Si+Si and Pb+Pb collisions (color online). } 
\label{centrality1}
\end{figure}

Fig. \ref{centrality1} shows the dependence of the width $\langle \Delta \eta \rangle$
of the BF on the mean number of wounded nucleons $\langle N_W \rangle$ (Table \ref{data_sets}). 
The results for p+p, C+C and Si+Si collisions are also included. The width decreases 
monotonically with $\langle N_W \rangle$. On the other hand the width of the BF from 
both HIJING and shuffled data does not show any clear dependence on centrality.

\begin{figure}[ht]
\begin{center}
\epsfig{angle=270,file=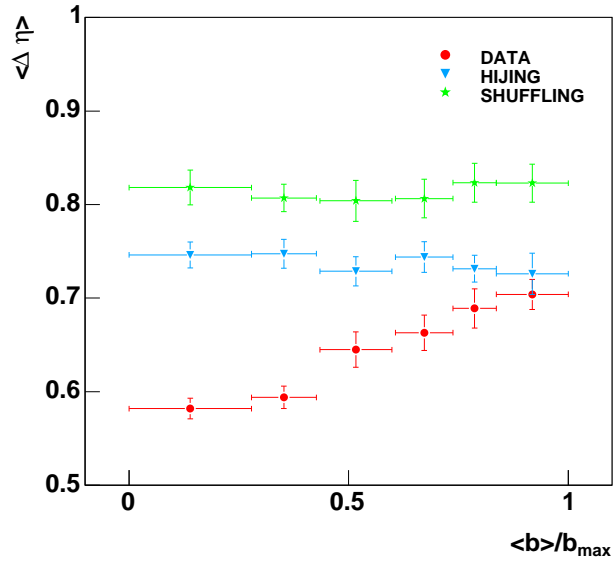,width=8cm}
\end{center}
\caption{The dependence of the BF 's width on the normalized
impact parameter $\langle b \rangle /b_{max}$ for Pb+Pb collisions (color online).}
\label{centrality2}
\end{figure}

\begin{figure}[ht]
\begin{center}
\epsfig{angle=270,file=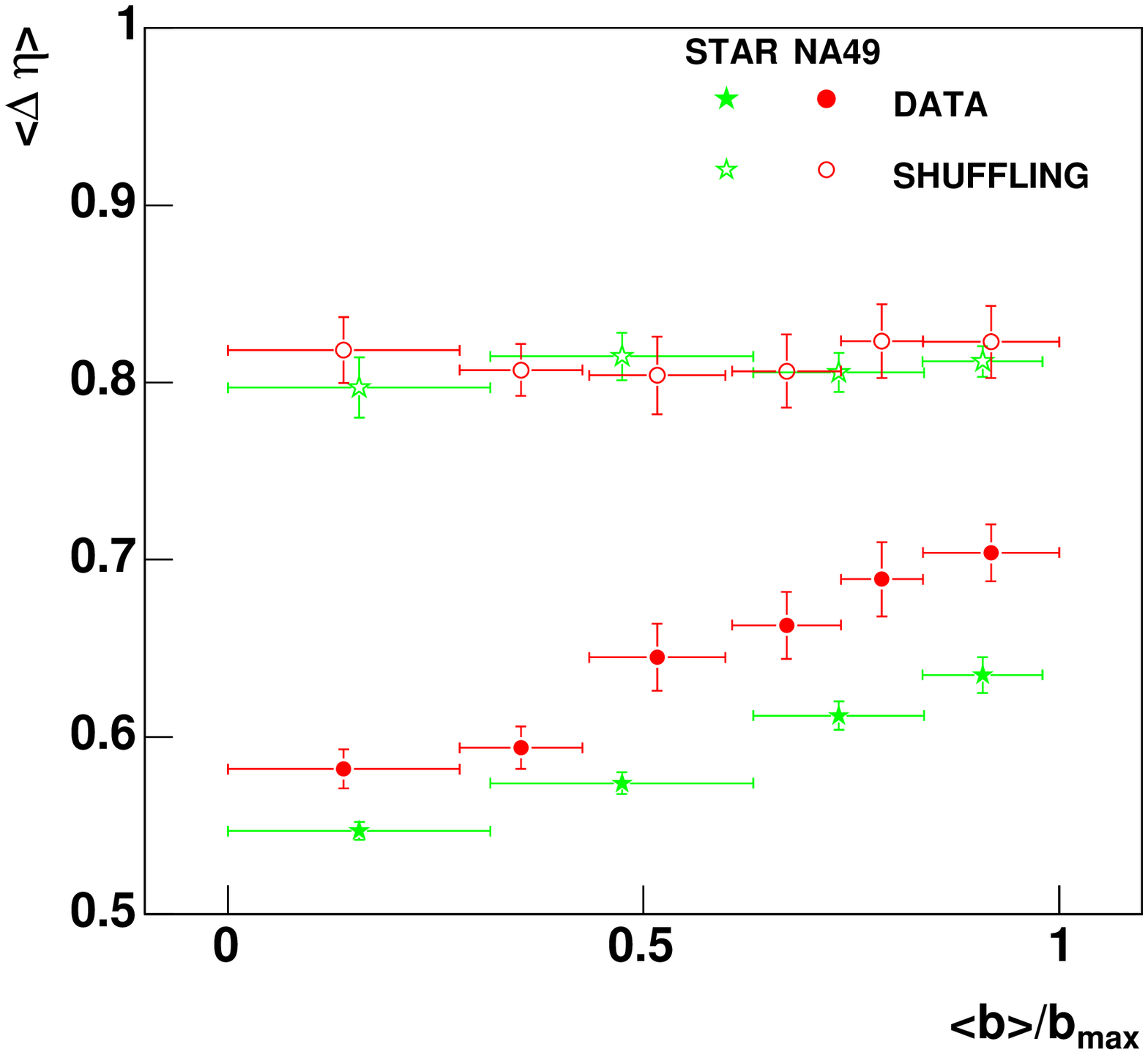,width=8cm}
\end{center}
\caption{The dependence of the BF 's width on the normalized
impact parameter $\langle b \rangle /b_{max}$, as measured by NA49 for Pb+Pb
collisions at $\sqrt{s_{NN}} = 17.2$ GeV and by STAR for Au+Au
collisions at $\sqrt{s_{NN}} = 130$ GeV (color online). } \label{STAR_NA49}
\end{figure}

Fig. \ref{centrality2} shows the dependence of $\langle \Delta \eta \rangle$ on the
normalized mean impact parameter $\langle b \rangle /b_{max}$. The values of the impact
parameter are listed in Table \ref{data_sets}. Once again the strong decrease of the width 
with increasing centrality of the collision is obvious. The results from a similar analysis 
performed for Au+Au collisions at $\sqrt{s_{NN}} = 130$ GeV by the STAR collaboration at 
RHIC \cite{STAR} are plotted in Fig. \ref{STAR_NA49}. The width of the BF decreases from 
peripheral to central collisions by $17 \pm 3\%$ for the NA49 data, whereas for the higher 
energy STAR data the corresponding decrease is of the order of $14 \pm 2\%$.

The narrowing of the BF compared to shuffled events is of similar magnitude in both 
experiments. The somewhat smaller difference between the widths for data and shuffled 
events for NA49 may be due to the incomplete azimuthal acceptance.

\vspace{0.5 cm}

The influence of the decay of resonances on the width of the BF was estimated using the 
HIJING event generator. We found that the BF width increases by about $4 \%$ when 
$\rho ^0$-meson decays are switched off. In the model the fraction of pions coming from 
$\rho ^0$ decays (about $19 \%$) is approximately independent of centrality. Therefore, 
the effect of $\rho ^0$ decay can not explain the strong system size and centrality 
dependence of the width of the BF that we observe in our experimental data.

The measured narrowing of the BF is qualitatively consistent with the delayed hadronization 
scenario \cite{Pratt,STAR} of an initially deconfined phase. Several model calculations have 
been published which provide a more quantitative description \cite{Statistical,Resonance1,Resonance2,Bialas}. 
In particular, within models based on statistical hadronization and hydrodynamic expansion 
the width of the BF was found to decrease with increasing transverse collective velocity of 
the matter at freeze-out \cite{Statistical,Resonance1,Resonance2} and thus with the collision 
centrality. However, a quantitative description of the STAR data was possible only when the 
condition of global charge conservation (a single fireball model) \cite{Resonance1,Resonance2} 
was substituted by a stronger condition of charge conservation in sub volumes (a multi-fireball 
model) \cite{Statistical}. The quark coalescence model was applied to the hadronization of the 
deconfined phase in \cite{Bialas}. When including radial flow, good agreement with the STAR 
measurements was obtained also in this model calculation.

\section{Summary}

In this paper the first measurements of the Balance Function in p+p, C+C and Si+Si 
interactions as well as centrality selected Pb+Pb collisions at $\sqrt{s_{NN}} = 17.2$ 
GeV (the top SPS energy) are presented. 

The width of the BF decreases monotonically with increasing system size (from minimum 
bias p+p to central Pb+Pb collisions) by $24 \pm 2\%$ and with increasing centrality of 
Pb+Pb collisions (from peripheral to central collisions) by $17 \pm 3\%$. A similar decrease, 
of the order of $14 \pm 2\%$, with centrality in Au+Au collisions was measured by STAR 
at $\sqrt{s_{NN}} = 130$ GeV. Thus the narrowing of the BF seems to be nearly energy 
independent from the top SPS to RHIC energies.

Events from the string-hadronic HIJING model as well as shuffled events retaining only 
correlations from global charge conservation do not show any significant decrease of 
the BF width with increasing system size and centrality in nucleus-nucleus collisions. 
On the other hand, results from central Pb+Pb reactions at top SPS and Au+Au reactions
at RHIC energies show a narrowing of the BF which suggests a delayed hadronization of 
the produced matter. For a more quantitative description of the data model calculations 
have to include the effect of transverse flow of the matter at freeze-out.

The energy dependence of the BF in the SPS range will be addressed in a future publication.

\vspace{2 cm}

\noindent
{\bf{Acknowledgments}}

\noindent This work was supported by the University of Athens/Special 
account for research grants, the US Department of EnergyGrant DE-FG03-97ER41020/A000,
the Bundesministerium fur Bildung und Forschung, Germany,
the Polish State Committee for Scientific Research (2 P03B 130 23, SPB/CERN/P-03/Dz 446/2002-2004,
2 P03B 04123), the Hungarian Scientific Research Foundation (T032648, T032293, T043514),
the Hungarian National Science Foundation, OTKA, (F034707),
the Polish-German Foundation, and the Korea Research Foundation Grant (KRF-2003-070-C00015).

\newpage

\begin{table}[ht]
\begin{center}
\begin{tabular}{|c|c|c|c|c|c|}
\hline  Interaction & Number of events & $E_0$ range (GeV) & $\langle N_{W} \rangle$ & $\langle b \rangle$ [fm] \\
\hline \hline p+p & 1M & & 2 & \cr

\hline C+C & 100K & & 14 & 1.9 \cr

\hline Si+Si & 100K & & 37 & 2.0 \cr

\hline Pb+Pb(6) & 300K & 29340 - 40000 & 42  & 11.5 \\ 

\hline Pb+Pb(5) & 110K & 26080 - 29340 & 88 & 9.6 \\ 

\hline Pb+Pb(4) & 88K & 21190 - 26080 & 134 & 8.3 \\ 

\hline Pb+Pb(3) & 75K & 14670 - 21190 & 204 & 6.5 \\ 

\hline Pb+Pb(2) & 100K & 9250 - 14670 & 281 & 4.6 \\ 

\hline Pb+Pb(1) & 100K & 0 - 9250 & 352 & 2.4 \\ 
\hline
\end{tabular}
\end{center}
\caption{Systems and centrality classes used in this analysis. Listed for p+p, C+C,
Si+Si and six centralities of Pb+Pb collisions at 158\emph{A} GeV are the range of 
the VCAL energy $E_0$, the mean number $\langle N_{W} \rangle$ of wounded nucleons 
and the mean value of the impact parameter.}
\label{data_sets}
\end{table}

\begin{table}[ht]
\begin{center}
\begin{tabular}{|c|c|c|c|c|c|c|c|c|}
\hline  Interaction&Year&Pol.&$V_{x0}$[cm]&$\Delta x$[cm]&$V_{y0}$[cm]&$\Delta y$[cm]&$V_{z0}$[cm]&$\Delta z$[cm]\\
\hline \hline p+p &2000&+& 0.0 & 1.0 & 0.0 & 1.0 & -580.0 & 5.0 \cr

\hline C+C &1998&+& 0.0 & 1.0 & 0.0 & 1.0 & -579.1 & 2.0 \cr

\hline Si+Si &1998&+& 0.0 & 0.3 & 0.0 & 0.5 & -579.5 & 1.0 \cr

\hline Pb+Pb (m.b.) &1996&-& 0.0 & 0.1 & 0.0 & 0.1 & -578.9 & 0.4 \cr

\hline Pb+Pb (m.b.) &1996&+& -0.05 & 0.1 & 0.05 & 0.1 & -578.9 & 0.4 \cr

\hline Pb+Pb (m.b.) &2000&+& 0.0 & 0.1 & 0.0 & 0.1 & -581.2 & 0.4 \cr

\hline Pb+Pb (cen.) &1996&+& 0.0 & 0.1 & 0.0 & 0.1 & -578.9 & 0.4 \cr
\hline
\end{tabular}
\end{center}
\caption{The different data sets used in the analysis. Listed for p+p, C+C,
Si+Si and different sets of Pb+Pb collisions at $\sqrt{s_{NN}} = 17.2$ GeV 
are the data taking period (Year), the field polarity (Pol.) as well as event 
selection cuts (see text for details).}
\label{sets}
\end{table}
\vspace{4 cm}

\begin{table}[ht]
\begin{center}
\begin{tabular}{c||c|c|c|c}
$y$& $A$ [c/GeV]  & $B$ [GeV/c] & $C$ $[deg \cdot (GeV/c)^{1/6}] $& $D$ [deg]  \\  \hline \hline
 -0.6  & 0   & 0   & 0  & 0\\
  -0.4  &0   & -1  & 63 & -8\\
 -0.2  & 0   & 0   & 57 & -10\\
 0.0  &  0   & 0.09& 63 & -13\\
 0.2  &  0   & 0.08& 67 & -4\\
 0.4  &  -7  & 0.08& 65 & -3\\
 0.6  &  0   & 0.05& 27 & 0\\
 0.8  &  0   & 0   & 35 & 0\\
 1.0  &  0   & 0.1 & 41 & 0\\
 1.2  &  0.34& 0.43& 109& 0\\
 1.4  &  0.36& 0.43& 100& 0\\
 1.6  &  0.55& 0.4 & 100& 0\\
 1.8  &  0.6 & 0.4 & 88 & 0\\
 2.0  &  0.61& 0.35& 73 & 0\\
 2.2  &  0.73& 0.34& 55 & 0\\
 2.4  &  1.7 & 0.28& 60 & 0\\
 2.6  &  2.8 & 0.25& 60 & 0\\
 2.8  &  5   & 0.2 & 57 & 0\\
 3.0  &  7   & 0.15& 60 & 0\\
 3.2  &  7   & 0.1 & 70 & 0\\
\end{tabular}
\end{center}
\vspace{2 cm}
\caption{Values of the parameters $A$, $B$, $C$ and $D$ of the acceptance
curves (Eq. \ref{Acc_Jac}). In the first column, the lower limit of the rapidity interval, 
$y$ is given. $y$ is calculated in the center of mass system assuming pion mass for all particles.} \label{abcd}
\end{table}

\newpage

\begin{table}[!h]
\begin{center}
\begin{tabular}{|c|c|c|c|}
\hline  Interaction & $\langle \Delta \eta \rangle$ (Data set 1)& $\langle \Delta \eta \rangle$ (Data set 2) & $\langle \Delta \eta \rangle$ (Data set 3)  \\
\hline \hline p+p & - & - & $0.767 \pm 0.007 $ \\ 

\hline C+C & - & - & $0.721 \pm 0.015 $ \\ 

\hline Si+Si & - & - & $0.698 \pm 0.011$ \\ 

\hline Pb+Pb(Veto 6) & $0.698 \pm 0.022  $ & $0.695 \pm 0.019  $ & $0.704 \pm 0.016  $ \\ 

\hline Pb+Pb(Veto 5) & $0.695 \pm 0.022  $ & $0.700 \pm 0.021  $ & $0.689 \pm 0.021  $ \\ 

\hline Pb+Pb(Veto 4) & $0.653 \pm 0.021  $ & $0.672 \pm 0.019  $ & $0.663 \pm 0.019  $ \\ 

\hline Pb+Pb(Veto 3) & $0.642 \pm 0.021  $ & $0.661 \pm 0.018  $ & $0.645 \pm 0.019  $ \\ 

\hline Pb+Pb(Veto 2) & $0.594 \pm 0.012  $ & - & - \\ 

\hline Pb+Pb(Veto 1) & $0.582 \pm 0.011  $ & - & - \\ 
\hline
\end{tabular}
\end{center}
\caption{The width of the BF for the three different data sets described in the text. } \label{sigma_results}
\end{table}

\begin{table}[!h]
\begin{center}
\begin{tabular}{|c|c|c|}
\hline  Interaction &$\langle \Delta \eta \rangle$ (SHUFFLING) & $\langle \Delta \eta \rangle$ (HIJING)  \\
\hline \hline p+p & $0.784 \pm 0.007$ & $0.764 \pm 0.005$  \\ 

\hline C+C & $0.815 \pm 0.014 $ & $0.746 \pm 0.010$  \\ 

\hline Si+Si & $0.833 \pm 0.011 $ & $0.732 \pm 0.012$  \\ 

\hline Pb+Pb(Veto 6) & $0.823 \pm 0.020 $ & $0.726 \pm 0.022 $  \\ 

\hline Pb+Pb(Veto 5) & $0.823 \pm 0.021 $ & $0.732 \pm 0.014 $  \\ 

\hline Pb+Pb(Veto 4) & $0.806 \pm 0.021 $ & $0.744 \pm 0.016 $  \\ 

\hline Pb+Pb(Veto 3) & $0.804 \pm 0.022 $ & $0.729 \pm 0.016 $  \\ 

\hline Pb+Pb(Veto 2) & $0.807 \pm 0.015 $ & $0.747 \pm 0.015 $ \\ 

\hline Pb+Pb(Veto 1) & $0.818 \pm 0.018 $ & $0.746 \pm 0.014 $ \\ 
\hline
\end{tabular}
\end{center}
\caption{The width of the BF for the shuffled and HIJING data sets. } \label{sigma_results_shuf}
\end{table}

\newpage

\newpage

\newpage

\newpage

\begin{thebibliography}{99}
\bibitem{QGP} {J. Collins, M. Perry, \emph{Phys. Rev. Lett.} {\bf 34}, 1353 (1975).}
\bibitem{QM} {See for recent results: Proc. of Quark Matter 2004, \emph{J. Phys. G} 30 (2004).}
\bibitem{Alb94} {
T. Alber et al. (\emph{NA35 Collaboration}), \emph{Z. Phys.} {\bf C64} (1994) 195.

S. Afanasiev et al.(\emph{NA49 Collaboration}), \emph{Phys. Rev.} {\bf C66} (2002) 054902.

R. Stock, \emph{Nucl. Phys.} {\bf A661} (1999) 282c.

}
\bibitem{QM04} {
M. Gazdzicki, M. Gorenstein, \emph{Acta Phys. Polon.} B30, 2705 (1999)

M. Gazdzicki (\emph{NA49 Collaboration}) Proc. of Quark Matter 2004, \emph{J. Phys. G} 30 (2004), S701.}
\bibitem{Stock99} {

S. Jeon and V. Koch, \emph{Phys. Rev. Lett.}  {\bf 85} (2000) 2076.

M. Asakawa, U. Heinz and B. M\"{u}ller, \emph{Phys. Rev. Lett.}  {\bf 85} (2000) 2072.

H. Heiselberg and A. D. Jackson, \emph{Phys. Rev.} {\bf C63} (2001) 064904.

E. V. Shuryak and M. A. Stephanov, \emph{Phys. Rev.} {\bf C63} (2001) 064903.

H. Appelsh\"{a}user et al. (NA49 Collaboration), \emph{Phys. Lett.} {\bf B459} (1999) 679.

S. V. Afanasiev et al. (NA49 Collaboration), \emph{Phys. Rev. Lett.} {\bf 86} (2001) 1965.

G. Georgopoulos, P. Christakoglou, A. Petridis and M. Vassiliou, 
Proc. of "Correlations and Fluctuations 2002". 

T. Anticic et al. (NA49 Collaboration), arXiv:hep-ex/0311009.
}
\bibitem{Pratt} {S. A. Bass, P. Danielewicz and S. Pratt, \emph{Phys. Rev. Lett.}  {\bf 85} (2000) 2689.}
\bibitem{Drij} {D. Drijard et al, \emph{Nucl. Phys.}  {\bf B155} (1979) 269 \\
D. Drijard et al, \emph{Nucl. Phys.}  {\bf B166} (1980) 233\\
I.V. Ajinenko et al., \emph{Nucl. Phys.}  {\bf C43} (1989) 37.}
\bibitem{Aih} {R. Brandelik et al., \emph{Phys. Lett.}  {\bf B100} (1981) 357\\
M. Althoff et al., \emph{Z. Phys.}  {\bf C17} (1983) 5\\
H. Aihara et al., \emph{Phys. Rev. Lett.}  {\bf 53} (1984) 2199\\
H. Aihara et al., \emph{Phys. Rev. Lett.}  {\bf 57} (1986) 3140\\
P.D. Acton et al., \emph{Phys. Lett.}  {\bf B305} (1993) 415.}
\bibitem{STAR} {J. Adams et al., (STAR Collaboration) \emph{Phys. Rev. Lett.} {\bf 90} (2003) 172301.}

\bibitem{na49_nim} S. Afanasiev et al. (NA49 Collab.) Nucl. Instrum. Meth. {\bf A430}, (1999) 210 .
\bibitem{na49_global} L.S. Barnby et al. (NA49 Collab.), J.Phys. {\bf G25}, (1999) 469 .

\bibitem{Cooper} G. Cooper. Ph.D. thesis,  Department of Physics, University
of California, Berkeley (2000) LBNL-45467.
\bibitem{venus_ref} K. Werner, Phys. Rept. {\bf 232}, 87, (1993).
\bibitem{Jacek} {
J. Zaranek, \emph{Phys.Rev.C} {\bf 66}, 024905 (2002) 

J. Zaranek, Diploma Thesis (2003), University of Frankfurt.}
\bibitem{Distortions} {S. Pratt and S. Cheng, \emph{Phys.Rev.C} {\bf 68}, 014907 (2003) }.
\bibitem{Hijing} {Xin-Nian Wang, Miklos Gyulassy, \emph{Phys.Rev.D} {\bf 3501-3516} (1991) 44.}
\bibitem{Statistical} {S. Cheng et al., \emph{Phys.Rev.C} {\bf 69}, 054906 (2004)}.
\bibitem{Resonance1} {P. Bozek, W. Broniowski, W. Florkowski, nucl-th/0310062.}
\bibitem{Resonance2} {W. Florkowski, P. Bozek, W. Broniowski, nucl-th/0402028.}
\bibitem{Bialas} {A. Bialas, \emph{Phys. Lett. B} {\bf B579} (2004) 31.}


\end{thebibliography}
\end{document}